\pdfoutput=1
\documentclass[aps,prl,showpacs,twocolumn]{revtex4}
\usepackage{graphicx}
\usepackage{amssymb}
\usepackage{amsmath}
\usepackage{color}
\usepackage{dsfont}
\usepackage{paralist}
\usepackage{bm}
\usepackage{hyperref}
\usepackage{verbatim}

\def\cH{\hat{\cal H}}

\def\bk{{\bf k}}

\def\ha{\hat a}

\DeclareMathOperator{\sign}{sgn}

\begin{document}

\title{Disorder-driven transition in a chain with power-law hopping}

\author{M.~G{\"a}rttner$^{1,3}$, S.V.~Syzranov$^{2,3}$, A.M.~Rey$^{1,2,3}$, V.~Gurarie$^{2,3}$, L.~Radzihovsky$^{1,2,3}$}

\affiliation{
$^1$JILA, NIST, University of Colorado, Boulder, Colorado 80309, USA
\\
$^2$Physics Department, University of Colorado, Boulder, Colorado 80309, USA
\\
$^3$Center for Theory of Quantum Matter, University of Colorado, Boulder, Colorado 80309, USA
}

\begin{abstract}
  {We study a 1D 
  system with a power-law quasiparticle
  dispersion $\propto |k|^\alpha\sign k$	 in the presence of a short-range-correlated random potential and
  demonstrate that for $\alpha<1/2$ it 
  exhibits a disorder-driven 
  quantum phase transition with the critical properties similar to those of the localisation transition 
  near the edge of the band of a semiconductor in high dimensions, studied in Refs.~\cite{Syzranov:Weyl}
  and \cite{Syzranov:unconv}.
  Despite the absence of localisation in the considered 1D system, the disorder-driven transition
  manifests itself, for example,
  in a critical form of the disorder-averaged density of states.
  We confirm the existence of the transition by numerical simulations and
  find the critical exponents and the critical disorder strength as
  a function of $\alpha$.
  The proposed system thus
  presents a convenient platform for numerical studies of the recently predicted unconventional
  high-dimensional localisation effects
  and has the potential for experimental realisations in chains of
  ultracold atoms in optical traps.
  }
\end{abstract}



\date{\today}
\maketitle

It is generally believed that increasing disorder strength in a conducting
material in dimensions $d>2$ leads to the Anderson localisation transition\cite{Anderson:original}
with universal properties that depend only on the space dimensionality $d$.

However, as we have demonstrated recently\cite{Syzranov:Weyl,Syzranov:unconv}, in high dimensions $d > d_c$
(with $d_c= 4$ for conventional weakly doped semiconductors and $d_c=2$ for Dirac semimetals)
the phenomenology is significantly richer. 
Namely, a
material with a power-law quasiparticle spectrum $\xi_\bk\propto k^\alpha$ in the presence of a
short-range random potential
in high dimensions $d>2\alpha$ 
exhibits an unconventional disorder-driven quantum phase transition in the bottom of the band, that
lies in a universality class distinct from the Anderson transition\cite{Syzranov:unconv,Syzranov:Weyl}.
Almost 30 years ago\cite{DotsenkoDotsenko,Fradkin1,Fradkin2} the existence of such a
transition was suggested for the specific case $\alpha=1, d=3$ for 3D Dirac materials
that have been later extensively studied in the literature\cite{Goswami:TIRG,Syzranov:Weyl,Brouwer:WSMcond,Moon:RG,Herbut,Syzranov:unconv,Pixley:2trans}
establishing a consensus for the existence of this novel
transition in 3D Dirac semimetals. 
Recently\cite{Syzranov:unconv,Syzranov:Weyl} we have shown the existence 
of such an unconventional disorder-driven transition and studied its properties for arbitrary $\alpha$ and $d$,
such that $d>2\alpha$, demonstrating that it is a generic property of high
dimensions and is not specific just to Dirac semimetals. 

For materials in $d>2$ dimensions in the symmetry classes that allow for localisation, this transition coincides with
the localisation transition
for the states in the bottom of the band\cite{Syzranov:unconv}.
However, it exists even if all states are always localised (e.g., in $d<2$ dimensions) or
if localisation is disallowed
by symmetry [e.g., in 3D Weyl semimetals (WSMs) with sufficiently
smooth disorder\cite{Wan:WeylProp,RyuLudwig:classification}]
and manifests itself, for example, in the critical behaviour of the density of
states and conductivity.

So far such unconventional disorder-driven 
transition has yet to be observed experimentally. 
Perhaps the main obstacle in 3D Dirac materials is the long-range Coulomb nature of quenched
disorder distinct from short-range random potential required to observe the critical behaviour
of the conductivity\cite{Syzranov:Weyl} (although the transition in the density of states is
still observable
in the presence of Coulomb impurities).
Another possible platform for studying high-dimensional localisation phenomena is
periodically-kicked quantum-rotor systems, that can be mapped\cite{Casati:mapping,Grempel:mapping}
onto high-dimensional
semiconductors with quadratic spectra. Such systems have been used to
simulate 1D\cite{Moor:rotorrealisation}, 2D\cite{Delande:rotorrealisation2D}, and 3D\cite{Chabe:rotorrealisation,Delande:rotorrealisation}
Anderson localisation, but the case of higher dimensions still remains to be realised.

Numerical simulations in high dimensions may be extremely 
demanding in terms of computing power. For instance, the quadratic spectrum of long-wave excitations,
generic for lattice models with short-range hopping and inversion symmetry, corresponds to $d_c=4$
and thus requires simulations
in $d\geq 5$ dimensions, with the number of sites growing rapidly
$\propto L^d$ as a function of the linear size $L$ of the system.

In this paper we suggest and study a new playground for unconventional
disorder-driven transitions, which
is rather convenient for numerical simulations and is also currently accessible for experiments:
1D systems with long-range hopping. 

Because the concept of high dimensions $d$ is
defined\cite{Syzranov:Weyl,Syzranov:unconv}
relative to the quasiparticle spectrum, the physics of high-dimensional
disorder-driven transitions
can be observed in any dimension $d$ by appropriately designing the inter-site hopping;
e.g., realising
the quasiparticle spectra $\propto |k|^\alpha$ with $\alpha<d/2$.
For instance, the spectrum $\propto |k|^\alpha$ with $\alpha<1$ in $d=1$ requires 
the inter-site hopping $\propto r^{-1-\alpha}$ which has already been realised
in 1D chains\cite{Monroe:longrange,Islam:longrange} and 2D arrays\cite{Bollinger:longrange}
of ultracold trapped ions.

Utilising fractional $\alpha<1/2$ in 1D systems also allows one to compare
the properties of the unconventional disorder-driven transition for $|\alpha-1/2|\ll1$ with theoretical
predictions\cite{Syzranov:Weyl,Syzranov:unconv} based on the RG approaches controlled by the small
parameter $\varepsilon=d-2\alpha$.

{\it Model.} In this paper we focus on a chiral (lacking reflection
symmetry)
1D system described by the Hamiltonian
\begin{equation}
	\cH=|k|^\alpha\sign k+U(x),
	\label{H}
\end{equation}
where $k$ and $x$ are momentum and coordinate and $U(x)$ is a short-range-correlated random potential.

We emphasise that ``chiral system'' hereinafter
means a system without reflection symmetry of the quasiparticle dispersion
($k\rightarrow -k$)
and should not
be confused with the concept of a system in a chiral symmetry class\cite{Zirnbauer:classes,AltlandZirnbauer}.

In some sense, such a system is a 1D analogue of a 3D Weyl semimetal. Indeed, the quasiparticle spectrum
consists of two bands: the conduction band ($k>0$) and the valence band ($k<0$),
touching at the node $k=0$. Quasiparticles in such a system
cannot be localised due to the absence of backscattering; the velocity $v(k)=\alpha |k|^{\alpha-1}$
never changes sign. In principle, the spectrum of a realistic 1D system on a lattice contains 
an equal number of branches with left- and right- movers, due to the continuity and periodicity
of the velocity $v(k)$ (analogously, Weyl semimetal has an even number of Weyl points\cite{Wan:WeylProp}).
However, for sufficiently smooth disorder elastic scattering of long-wavelength 
quasiparticles to states far from the node ($k=0$) can be neglected and the quasiparticle
dynamics near the node can be described by the model \eqref{H}.

The quenched disorder potential $U(x)$ with zero mean and a symmetric
distribution function preservers the $E\rightarrow-E$ symmetry of the quasiparticle spectrum and
the density of states $\rho(E)$ (with the energy $E$ measured from the node),
which makes the chiral system particularly convenient for numerical studies of the disorder-driven
transition near the node.
In contrast, in a system with a single band (corresponding, e.g., to the spectrum $|k|^\alpha$) quenched
disorder generically leads to the renormalisation of the band edge and to the formation of Lifshitz tails
below the band\cite{Syzranov:unconv}, making it hard to define and to
identify numerically the renormalised edge of the band.

{\it Disorder-driven transition.}
For $\alpha$ slightly smaller than $1/2$ the effects of disorder
can be analysed using a renormalisation-group (RG) approach, controlled by the
small parameter $\varepsilon=2\alpha-1$. This RG, previously applied to Dirac
materials\cite{LudwigFisher,Nersesyan:dwave,Goswami:TIRG,AleinerEfetov,Syzranov:Weyl,OstrovskyGornyMirlin,
Moon:RG,Syzranov:unconv} and to high-dimensional semiconductors in the orthogonal symmetry class\cite{Syzranov:unconv},
repeatedly removes the highest momenta from the system, renormalising its properties at lower momenta.
Depending on whether or not the characteristic amplitude $W$ of the random potential exceeds a critical value $W_c$,
the dimensionless strength of disorder $\gamma\sim 1/(k\ell)$ flows to larger or smaller values under the RG, where
$\ell$ is the (flowing) mean free path. Such behaviour of the renormalised disorder strength signifies 
a phase transition between the weak- and strong-disorder phases at $E=0$. In this paper we verify numerically
that such a transition persists at all $\alpha<1/2$.

The behaviour of the low-energy density of states near a critical point has
the generic scaling form (first proposed for 3D Dirac materials in Ref.~\cite{Herbut})
\begin{equation}
	\rho(E,W)=E^{\frac{d}{z}-1}\Phi\left[(W-W_c)/E^\frac{1}{z\nu}\right],
	\label{RhoGeneral}
\end{equation}
with $\nu$ and $z$ being the correlation-length and dynamical critical exponents.
In the weak-disorder phase ($W<W_c$) the density of states has the same energy dependency as free
quasiparticles $\rho(E,W)\propto(W_c-W)^{-\nu\left(\frac{z}{\alpha}-1\right)}E^{\frac{1}{\alpha}-1}$,
while for strong disorder ($W>W_c$) the density of states is smeared and thus energy independent:
$\rho(E,W)\propto(W-W_c)^{(1-z)\nu}$. For finite energy $E$ these two regimes are separated by
a critical region near $W=W_c$ with $\rho(E)\propto E^{1/z-1}$.

The RG analysis similar to that of Refs.~\cite{Syzranov:Weyl} and \cite{Syzranov:unconv} for
the model (\ref{H}) in
the one-loop approximation with the small parameter $\varepsilon=2\alpha-1$ yields
\begin{align}
	\nu & =  ({1-2\alpha})^{-1},
	\label{nu}
	\\
	z & =  {1}/{2}.
	\label{z}
\end{align}
We emphasise that the dynamical exponent $z$, Eq.~\eqref{z},
is independent of $\alpha$ only in the first order in $\varepsilon$ and only
for the 1D chiral system under consideration. For arbitrary $\alpha$, not necessarily
close to $1/2$, the values of the critical exponents can be found numerically.

\begin{figure}[hb!]
	\centering
	\includegraphics[width=\columnwidth]{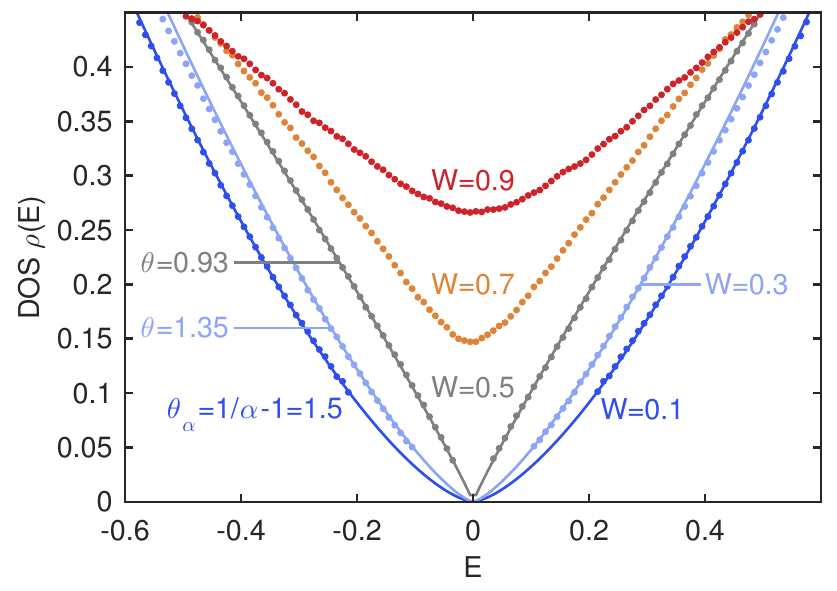}
	\caption{\label{DoSW}
	(Colour online)
	The disorder-averaged density of states $\rho(E)$ vs. energy $E$ for $\alpha=0.4$ and various disorder amplitudes $W$.
	For subcritical disorder strength, $W<W_c\approx0.5$, the density of states vanishes at $E=0$, whereas for stronger
	disorder, $W>W_c$, the density of states is finite for all energies. At the critical disorder strength,
	$W=W_c\approx0.5$, the density of states is linear in energy, in agreement with the analytical predictions
	based on one-loop RG calculations.}
\end{figure}

{\it Numerical results.} In what follows we present the results of the numerical simulations
that demonstrate the existence of the above described disorder-driven transition for $\alpha<1/2$
and its absence for $\alpha>1/2$. We also analyse the critical behaviour near the transition
and obtain numerically the values of the critical exponents.

To simulate the model \eqref{H}, we use its lattice version
\begin{equation}
	\cH=\sum_{x,x^\prime}J_{xx^\prime}\ha^\dagger_x\ha_{x^\prime}+\sum_x U_x\ha^\dagger_x\ha_x,
	\label{Hlattice}
\end{equation}
of finite size $N$ with periodic boundary conditions, where distances (momenta)
are measured in (inverse) lattice spacings;
$J_{x,x^\prime}=J_{x-x^\prime}=\sum_k e^{ik(x-x^\prime)} |k|^\alpha\sign k$ is the inter-site hopping
element (that we find numerically), for long distances $|x-x^\prime|\gg1$ given by the
odd power-law function\footnote{
We emphasise, that the model \eqref{Hlattice} with power-law hopping
$J_{xx^\prime}\propto|x-x^\prime|^{-(1+\alpha)}\sign(x-x^\prime)$, that we consider, should not be
confused with the power-law random banded matrix model\cite{MirlinEvers}, characterised by random Gaussian hopping
with the power-law dependence on distance.
}
\begin{equation}
	J_{x-x^\prime}=i\:
	\frac{\sign(x-x^\prime)}{|x-x^\prime|^{1+\alpha}}\:\frac{\Gamma(1+\alpha)}{2\pi}
	\:\sin\left[\frac{\pi}{2}(1+\alpha)\right];
\end{equation}
and $U_x$ is the random disorder potential, uncorrelated on different sites and described by the Gaussian
on-site
distribution with standard deviation $W$,
$P(U_x)=(W\sqrt{2\pi})^{-1}\exp[-U_x^2/(2W^2)]$. 
For the on-site-correlated disorder under consideration 
the ultraviolet momentum cutoff\cite{Syzranov:unconv}
$K_0\sim 1$
is determined by the lattice spacing. 

\begin{widetext}

\begin{figure}[h] 
\includegraphics[width=.9\textwidth]{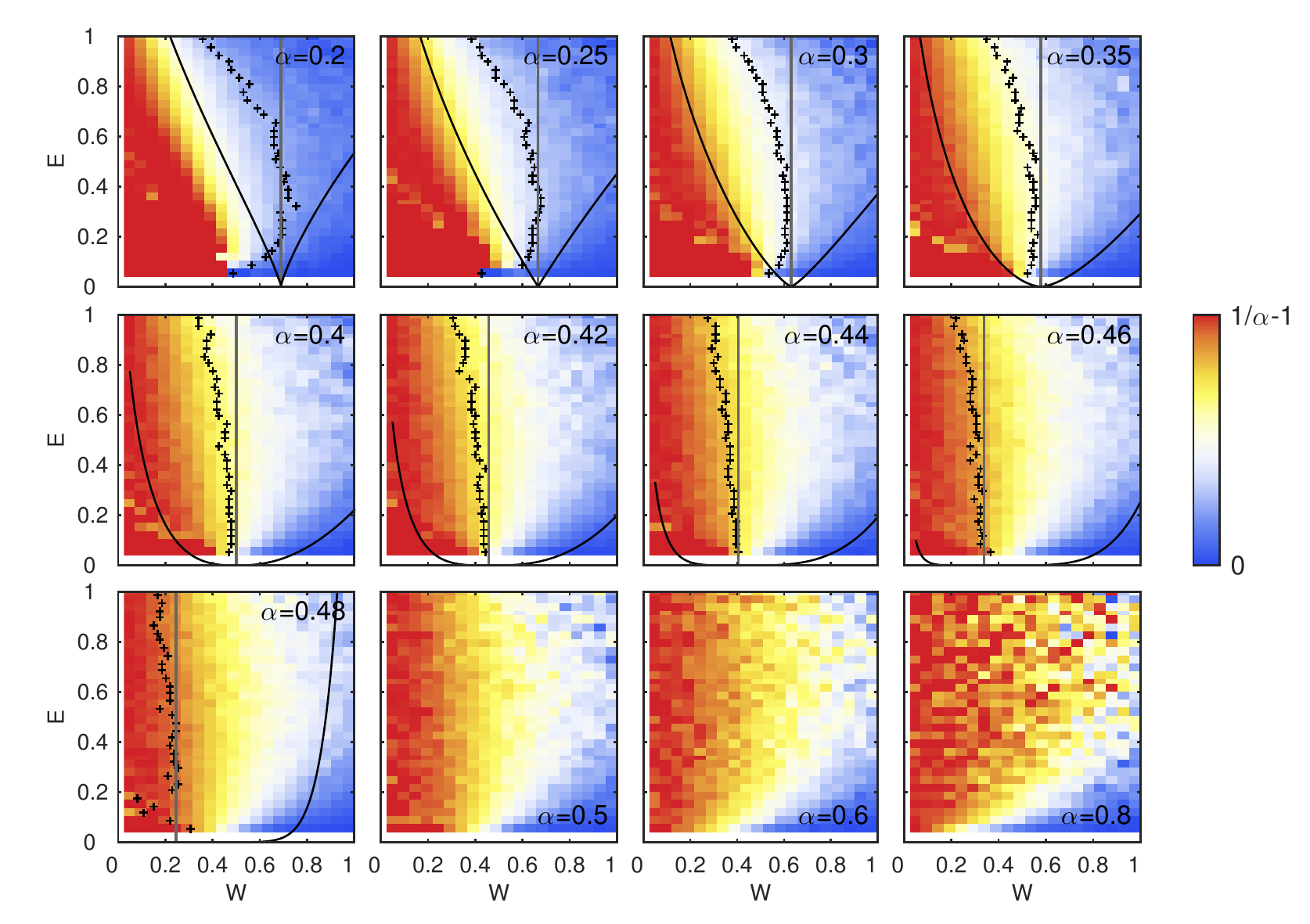}
\caption{(Colour online)
\label{Results}
Energy (E) vs. disorder amplitude (W) diagram for the low-energy density of states
in a 1D chiral system with the quasiparticle spectrum $\xi_k=a |k|^\alpha\sign k$.
The colour shows the exponent $\theta$ of the density of states $\rho(E)\propto E^\theta$ [see the colourbar
and Eq.~\ref{theta}].
For $\alpha<1/2$ the density of states has a critical point ($E=0$, $W=W_c$) that separates
 the weak-disorder
($\theta_\alpha=\alpha^{-1}-1$) and strong-disorder
($\theta=0$) phases at $E=0$.
The transition disappears for $\alpha>1/2$.
The grey vertical line shows the
analytical value of the critical disorder amplitude obtained assuming $0<1-2\alpha\ll1$.
The black crosses show the isoline $\theta=1$.
The black solid lines show the theoretical crossover energies $E=(|W-W_c|/W_c)^{z\nu}$
between the critical region ($\theta=1/z-1$) near $W\approx W_c$
and the weak- ($\theta_\alpha=\alpha^{-1}-1$) and strong- ($\theta=0$) disorder phases, with the exponents
$\nu$ and $z$ given by Eqs.~(\ref{nu}) and (\ref{z}).
}
\end{figure}

\end{widetext}

For each value of $\alpha$ and disorder strength 
we use the exact diagonalisation method to obtain
the spectrum of the system for 100 disorder realisations on a lattice with $N=4000$ sites
and find the parameter
\begin{equation}
	\theta(E,W)=\frac{\partial\ln\rho(E,W)}{\partial\ln E}
	\label{theta}
\end{equation}
as a function of the random potential amplitude $W$ and energy $E$, with the results
summarised in Fig.~\ref{Results}.

For sufficiently small $\alpha$ we observe a disorder-driven quantum phase transition;
the zero-energy density of states vanishes
\footnote{We emphasise, that exponentially rare strong fluctuations of the disorder potential
may lead to a finite density of states\cite{Nandkishore:rare},\cite{Syzranov:unconv}
even for $W<W_c$, similarly to the formation of Lifshitz tails
in semiconductors\cite{Lifshitz:tail,ZittartzLanger,HalperinLax,LifshitzGredeskulPastur}.
We believe, the accuracy of our numerical simulations is insufficient to observe
such rare-region contributions, if they exist for the system under consideration.}
for disorder amplitudes smaller
than a critical value, $W<W_c$, and has a finite value otherwise [the dependency $\rho(E)$
for $\alpha=0.4$ is shown in Fig.~\ref{DoSW}].

For
sufficiently small $\alpha$
the density of states
displays three regions
with qualitatively different behaviours, that touch
at a critical point $E=0$, $W=W_c$, Fig.~\ref{Results}: (i) for low energies and disorder strengths 
the density of states has the energy dependency $\rho(E)\propto E^{1/\alpha-1}$ of free quasiparticles with the
spectrum $\propto |k|^\alpha\sign k$ ($\theta=1/\alpha-1$, red colour in Fig.~\ref{Results});
(ii) for low energies and sufficiently strong disorder 
the density of states is constant, $\theta=0$, 
(blue colour in Fig.~\ref{Results});
(iii) near the critical disorder strength, $W=W_c$, there is an intermediate critical region
with an almost constant intermediate value of $\theta$. These results confirm the existence of the
disorder-driven phase transition for sufficiently small $\alpha$.

In order to accurately verify that the criticality disappears for $\alpha>1/2$,
we utilise the theoretical predictions of the one-loop RG analysis for the critical properties of the transition
near $\alpha=1/2$.
Indeed, such RG analysis is controlled by the small parameter $\varepsilon=2\alpha-1$ and thus
becomes exact when this parameter vanishes. 

The one-loop RG analysis predicts the critical
exponents (\ref{nu}) and (\ref{z}) and the density of states 
$\rho(E)\propto E^{1/z-1}=E$ at the critical disorder strength ($W=W_c$), corresponding to $\theta=1$,
which can be used to accurately identify the critical point. 
In Fig.~\ref{Results} 
the points with $\theta=1$ are shown by black crosses that form a line
which contains the critical point and is vertical at low energies.

The respective value of the critical disorder amplitude $W_c$ matches well the
result (shown by the grey vertical solid line in Fig.~\ref{Results})
\begin{equation}
	W_c=\left[\pi (1-2\alpha)/2\right]^\frac{1}{2} K_0^{\alpha-\frac{1}{2}}
\end{equation}
of the one-loop RG calculation that we obtain under the assumption $0<1-2\alpha\ll1$ with the ultraviolett momentum cutoff $K_0=\pi$
(for details of the scheme of the perturbative RG calculation see Ref.~\cite{Syzranov:unconv}),
even for $\alpha$ significantly below $0.5$.

These results demonstrate the existence of the criticality of the density of states
for $\alpha<0.5$ and its disappearance for $\alpha>0.5$. For all values of $\alpha$ in the interval
$0.2\ldots0.5$ the critical
properties of the transition are well described by the results of the one-loop perturbative RG analysis.

{\it Conclusion and outlook.} In summary, we have demonstrated that a chiral 1D system with the quasiparticle
spectrum $|k|^\alpha\sign k$ with $\alpha<1/2$ displays the phenomenology of a
high-dimensional disorder-driven phase transition. 
Although all the states in the proposed system are delocalised, it exhibits a disorder-driven
transition that manifests itself in the density of states and is analogous to the localisation
transition near the edge of the band of a high-dimensional semiconductor. In terms of its symmetries and
the critical behaviour of observables, the system under consideration presents a 1D analogue
of a 3D Weyl semimetal. The numerical values of the critical exponents and the critical disorder
strength are well described by the results
of a one-loop perturbative RG calculation.
Such a system presents a convenient platform for studying high-dimensional localisation physics
and can be used to further investigate strong-disorder conduction in semimetals with delocalised states,  
the interplay of disorder with interactions, effects of various disorder symmetries on the transition, etc.

We emphasise, that the unconventional disorder-driven transition we studied is not specific to
chiral 1D systems and can be observed (with different critical exponents)
in any 1D chain with sufficiently long-range hopping of the excitations,
e.g., corresponding to the even dispersion $\xi_k\propto |k|^\alpha$ with $\alpha<1/2$.
However, for systems with non-odd dispersions $\xi_k$
the transition may be harder to observe numerically and experimentally
due to the renormalisation of band edges or nodal points by disorder.

Implementing the specific 1D chiral model in experiments still remains a future research direction.
Natural candidates are chains of trapped ions, since in those systems a power-law 
excitation spectrum $\propto |k|^\alpha$ with tunable $0<\alpha<1.5$ has already 
been demonstrated\cite{Monroe:longrange,Islam:longrange}.
However, ways to generate chiral excitations in these chains still have to be investigated.
In principle, a non-chiral power-law spectrum $\propto |k|^\alpha$ is also suitable for the observations of the
high-dimensional localisation physics, but is less convenient for numerical simulations
and is more sensitive to finite-size effects, that we expect to obscure the disorder-driven transition
for non-chiral spectra for small numbers of ions ($<20$) used in the current experiments.
Another candidate for the observation of high-dimensional localisation physics is 2D arrays of ions
in Penning-trap experiments\cite{Bollinger:longrange}, where a tunable power-law spectrum has been demonstrated
for about 500 ions arranged in a triangular lattice.  We leave the analysis of finite-size effects
and possible realisations of chiral excitations in such systems for future studies.

{\it Acknowledgements.}
Our work was supported by the Alexander von Humboldt
Foundation through the Feodor Lynen Research Fellowship (SVS) and by the NSF grants
DMR-1001240 (LR and SVS),
DMR-1205303(VG and SVS), PHY-1211914 (VG, AMR, and SVS), and PHY-1125844 (SVS, MG, AMR).
LR also acknowledges support by the Simons Investigator award from the Simons Foundation.
AMR also acknowledges support from AFOSR, AFOSR-MURI, NIST and ARO individual investigator
awards.


\end{document}